# Gravitational Anomalies by HTC superconductors: a 1999 Theoretical Status Report.

## G. Modanese

## The Gravity Society - www.gravity.org

**Abstract -** In this report we summarize in an informal way the main advances made in the last 3 years and give a unified scheme of our theoretical work. This scheme aims at connecting in a consistent physical picture (by the introduction of some working hypotheses when necessary) the technical work published in several single articles. The part of our model concerning the purely gravitational aspects of the weak shielding phenomenon is almost complete; the part concerning the density distribution of the superconducting carriers in the HTC disks is still qualitative, also due to the very non-standard character of the experimental setup. The main points of our analysis are the following: coherent coupling between gravity and a Bose condensate; induced gravitational instability and "runaway" of the field, with modification of the static potential; density distribution of the superconducting charge carriers; energetic balance; effective equations for the field; existence of a threshold density.

| Main points of the research project | References |
|---|---|
| (1) Relevance and innovation of the effect. Impossibility of an explanation within General Relativity.. | SHI |
| (2) Logical core of our model: | |
| (a) Peculiar coupling betw. gravity and a Bose condensate $\rightarrow$ | SHI, LCC |
| (b) $\rightarrow$ Induced instability and "runaway" of the zero modes $\rightarrow$ | ISS; earlier in SHI, LCC |
| (c) $\rightarrow$ Constraint on the gravitational field and "tunneling". | GAU; earlier in SHI, LCC |
| (3) Density pattern of the superconducting carriers in the disk. | BOS |
| (4) Energetic balance and effective equations for the gravitational field | BOS |
| (5) Threshold density of the condensate | BOS |

The weak gravitational shielding effect by HTC superconductors discovered by E. Podkletnov and the transient gravitational anomalies observed by J. Schnurer open an incredibly wide research field, at the boundary between condensed matter physics and gravitation.

All the observed phenomenology strongly supports our proposed theoretical model, based on the idea that an instability of the gravitational field arises in the so-called "critical regions" of the superconducting charge carriers condensate.

The task of summarizing and connecting in a logical way all the conceptual and theoretical implications of these effects is hard, for two reasons at least.

(i) Up till now, we do not have any consistent theory which, starting from well known paradigms, leads in a purely technical way to an explanation of the experiments. It is impossible to "prove theoretically" the existence of the anomalies. Even though the starting point of our proposed model is entirely orthodox (quantum mechanics + general relativity + Ginzburg-Landau theory of superconductors), simplifications and unproven hypotheses are needed at several points.

This situation is not unusual in physics, particularly not for very complex systems. It happens quite often that subsequent developments justify "a posteriori" previous assumptions. The art of the theoretician consists also in making reasonable assumptions, which do not hurt any fundamental principle or prior knowledge and fit well with the experimental intuition.

(ii) There is a wide interest towards these new results, but specialists of different areas have different attitudes and questions. Some wonder how the effect is connected to the perturbative or non-perturbative dynamics of quantum gravity or its generalizations; others take for granted that there is an exotic effect at the basis of the anomalous coupling, and focus on phenomenological issues. We tried to address several "frequently asked questions" throughout the paper and the reader may notice by himself the variety of the question spectrum.

The organization of this article is a direct consequence of the two points above. On one hand, the fundamental logical structure of the theoretical model is stressed, with reference to the specific articles for more details; new assumptions and obscure points are discussed. On the other hand, the various issues are also listed as single items, like in a handbook, with some redundant repetitions or logical simplifications when necessary.

The main points of our analysis are the following.

**(1) Relevance and innovation of the effect. Impossibility to explain it within**

## classical General Relativity.

This is a necessary premise for any serious analysis of these effects. One should realize that the gravitational "anomalies" claimed by Podkletnov et al. are at least 10 orders of magnitude larger than those predicted by the General Relativity theory. For this reason, these results are hard to be accepted and understood. The scientific community reacted to these claims with a mixture of interest and scepticism.

It is important to notice, however, that no basic physical principles (like for instance symmetry or conservation laws) are contradicted by the claimed effects. The latter can be just ascribed to a peculiar dynamical behaviour of the gravitational field in certain circumstances. We do not see in the results of Podkletnov et al. any "revolutionary" content, such to justify a strong preconceived opposition. Furthermore, the effects have precise energetical limitations and can be observed only under certain conditions. See also on this point Section 4 and the remarks on the Equivalence Principle.

The following Points (2a), (2b), (2c) are the "logical core" of our theoretical model.

## (2a) Peculiar "coherent" coupling between gravity and a Bose condensate.

A Bose condensate has the property to be macroscopically coherent and is thus described by a classical field $\phi_0(x)$ - also called "order parameter" - whose squared module $|\phi_0(x)|^2$ is related to the condensate density.

This coherence property implies that the condensate contributes not only to the linear part of the gravitational lagrangian $L$, with standard coupling $\delta L^{(1)} = 8\pi h_{\mu\nu} T_{\mu\nu}(\phi_0)$, but also to the quadratic part. The contribution to the quadratic part affects locally the "potential" of the gravitational field, that is, that part of $L$ which does not depend on the field derivatives.

More precisely, the coherent coupling term is $\delta L_{\text{Coherent}}^{(2)} = \{\sqrt{\det[g(x)]}\}^{(2)} \mu^2(x)$, where $\mu^2(x)$ is a positive definite quantity defined as

(1) $\quad \mu^2(x) = \partial_\mu \phi_0^*(x) \, \partial_\nu \phi_0^*(x) + m^2 |\phi_0(x)|^2$

and $\{\sqrt{\det[g(x)]}\}^{(2)}$ denotes the quadratic part of $\sqrt{\det[g_{\mu\nu}(x) = \delta_{\mu\nu} + h_{\mu\nu}(x)]}$, which in terms of the field $h_{\mu\nu}(x)$ is negative definite and thus will be symbolically denoted in the following simply as "$-h^2$".

In the gravitational lagrangian there is already a term of this kind, namely $\{\sqrt{\det[g(x)]}\}^{(2)} (\Lambda/8\pi G)$, where $\Lambda$ is the very small negative intrinsic cosmological constant of spacetime. We define as "critical regions" those where $\mu^2(x) > |\Lambda|/8\pi G$, that is, those where the coherent coupling term is larger than the "threshold" term. Since the absolute value of $\phi_0(x)$ is related to the density of the Bose condensate, the critical regions are defined through eq. (1) by the condition that either the density of superconducting charge carriers or the gradient of this density are particularly large.

## (2b) Induced instability.

There exists a class of unstable modes of the pure Einstein action, called "zero modes", which have the same probability to occur as the $h=0$ configuration (flat space) and a higher probability than all other field configurations. (The probability is proportional to $\exp(iS[g]/h_{Planck})$, and for the zero-modes one has $S=0$, like for flat space; see the Discussion of Point 2a, "Quantum nature of the effect".)

Thus a gravitational field would always tend to "run away" towards these configurations and loose memory of its initial state, if it was not regulated by a small negative cosmological term, intrinsically present in nature. This term has the form $\delta L_{Cosm.}^{(2)}=+|\Lambda|h^2$, so it favours the $h=0$ configuration.

The coherent coupling $\delta L_{Coherent}^{(2)}=-\mu^2 h^2$ of the gravitational field to the condensate amounts to a local positive cosmological term (note that unfortunately the actual signs are opposite to the conventional denomination). This cancels the intrinsic stabilizing term $\delta L_{Cosm.}^{(2)}$ and leads to a local instability.

Therefore, while the regular coupling of gravity to incoherent matter produces a *response* - a gravitational field - approximately proportional to the strength $T_{\mu\nu}$ of the source, the coherent coupling induces an *instability* of the field.

Things go as if a potential well for the field was suddenly opened. The field runs away towards those configurations which are now preferred (compatibly with the global energetic balance - compare Section 4). The runaway stops at some finite strength of the field, where higher order terms in the lagrangian, which usually can be disregarded, come into play. This strength is not known since it depends on the non perturbative dynamics of the field, but very little on the initial conditions.

It is important to stress that not all the field modes become unstable and undergo this runaway; only the so-called zero modes have a high probability $\xi$ to do so (compare Fig. 1) and these modes are a little part of all possible configurations. Also for this reason (ratio of volumes in probability space, or "phase space"), the total probability of the process is quite small.

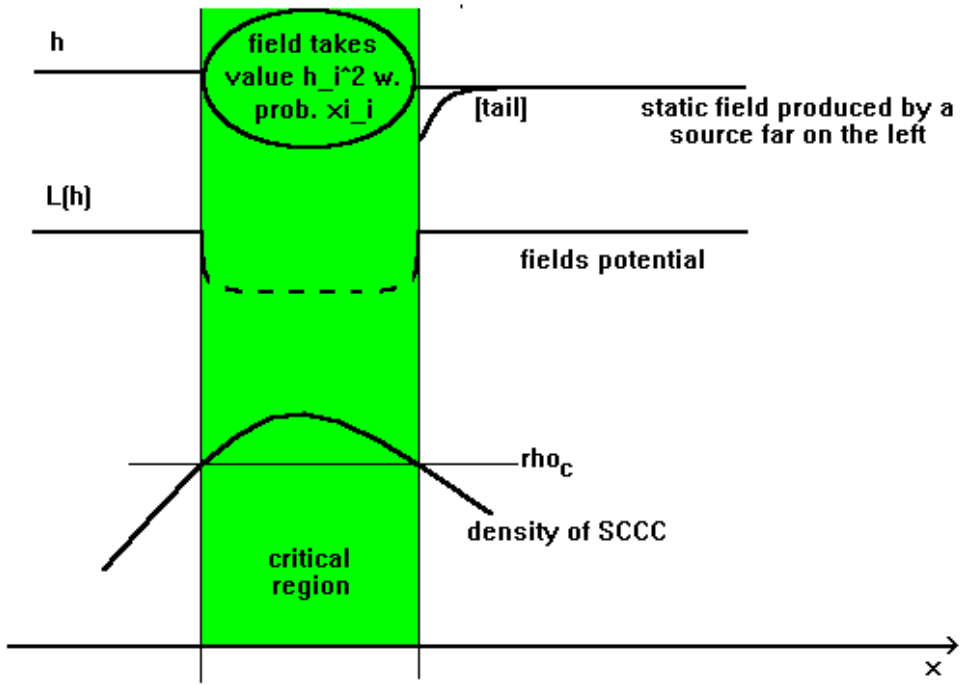

**Fig. 1**

This figure represents, for illustration purposes, the behavior across a critical region of:

*Bottom* - The density of superconducting charge carriers (SCCC). By definition, the critical region begins where the density of SCCC exceeds the threshold density $\rho_c$. We are not interested at this stage into the causes of this local density increase.

*Middle* - The potential part of the gravitational lagrangian $L(h)$, that is, the part which does not contain the field derivatives. This well-like shape of the potential is due to the coherent coupling and is responsible for the instability. The dotted line means that higher order terms in the lagrangian come into play and stop the instability - but we do not know exactly how and for which strength of $h$. Note that the instability arises quite sharply at the border of the critical region (we admit that a threshold value of $\Lambda$ exists - see Section 5).

*Top* - The static field produced by a far source, supposed to be on the left of the critical region. In the critical region the field "runs away" and takes strengths $h_i^2$ with probabilities $\xi_i$. This is a quantum process, and the strengths $h_i$ as well as their probabilities are non-observable parameters. Only the parameter $\gamma = \Sigma_i \, \xi_i \, h_i^2$ is actually observed, and must be inferred from the experimental data.

We did not represent in the graph the strength of $h$ within the critical region, just because only probabilities can be defined.

The computation in GAU allows us to conclude that the pre-existing static field is decreased in the critical region by a % amount proportional to $\gamma$ and to the thickness and width of the region (corresponding to a coherence domain of the condensate or to a part of it).

The field strength is most affected just outside the critical region (see 2c), then the original value is almost completely restored, apart from a little residual modification ("tail") which is the one actually observed.

## (2c) Constraint on the gravitational field and "tunneling".

What is the effect of the instability induced by the peculiar coherent coupling on a pre-existing static gravitational field, generated by a far source? Why does a weak shielding result?

This crucial point of our model was already studied qualitatively in Ref.s SHI, LCC, and has been clarified through an explicit calculation in GAU.

Intuitively, it is clear that the static field is affected by the presence of the potential well. As we mentioned earlier (compare also Fig.1) the field is forced to assume within the critical region, with certain probabilities, strengths independent of its original strength. Which is the residual effect of this constraint outside the critical region?

One could try some analogy with a tunneling effect in ordinary quantum mechanics. After doing a Fourier decomposition of the static field produced by the far source, one can identify those modes which have a large probability to be stopped by the potential well. It turns out, however, that it is impossible to carry on this qualitative analysis in a satisfying way. It is necessary to implement a rigorous "ab initio" calculation. This has been done in GAU, and the positive results give us further confidence in our model.

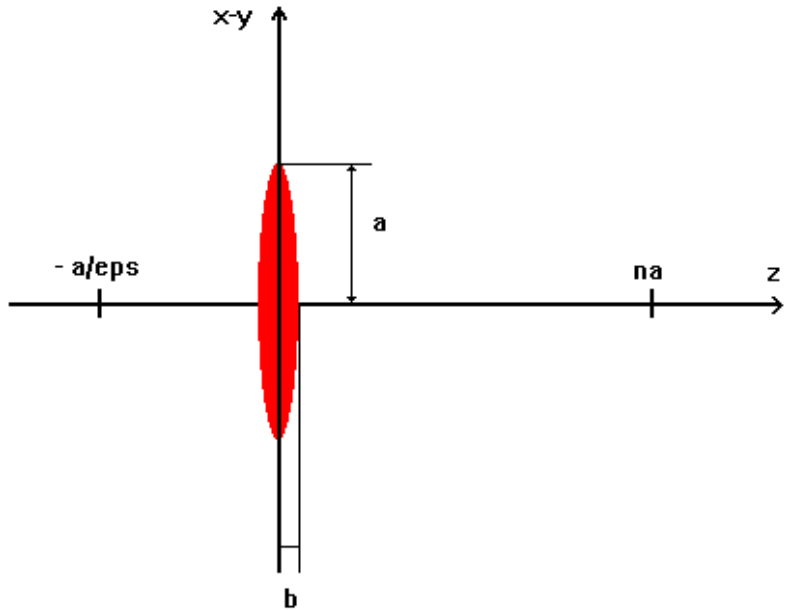

**Fig. 2** - Critical region with the shape of an ellipsoid.

In the special case of a massless field (like the gravitational field), we found that if the field, denoted by $f$, is forced to assume the values $f_i$ within a given region, with probabilities $\xi_i$, then the static potential of the interaction mediated by $f$ between two sources placed at the opposite sides of that region (Fig. 2) is decreased by the factor

(2)     $[1 - \gamma ab F(\varepsilon,\rho,n)]$,

where $\gamma = \Sigma_i \xi_i |f_i|^2$, $a$ and $b$ represent width and thickness of the region, $\varepsilon$ is the inverse of the distance of the first source in units of $a$, $\rho$ is the ratio $b/a$ and $n$ is the distance of the second source from the region, measured in units of $a$. Evaluating the function $F(n)$ for $\varepsilon$ and $\rho$ constant one obtains a decreasing exponential, with a non-zero asymptotic value (Fig. 3).

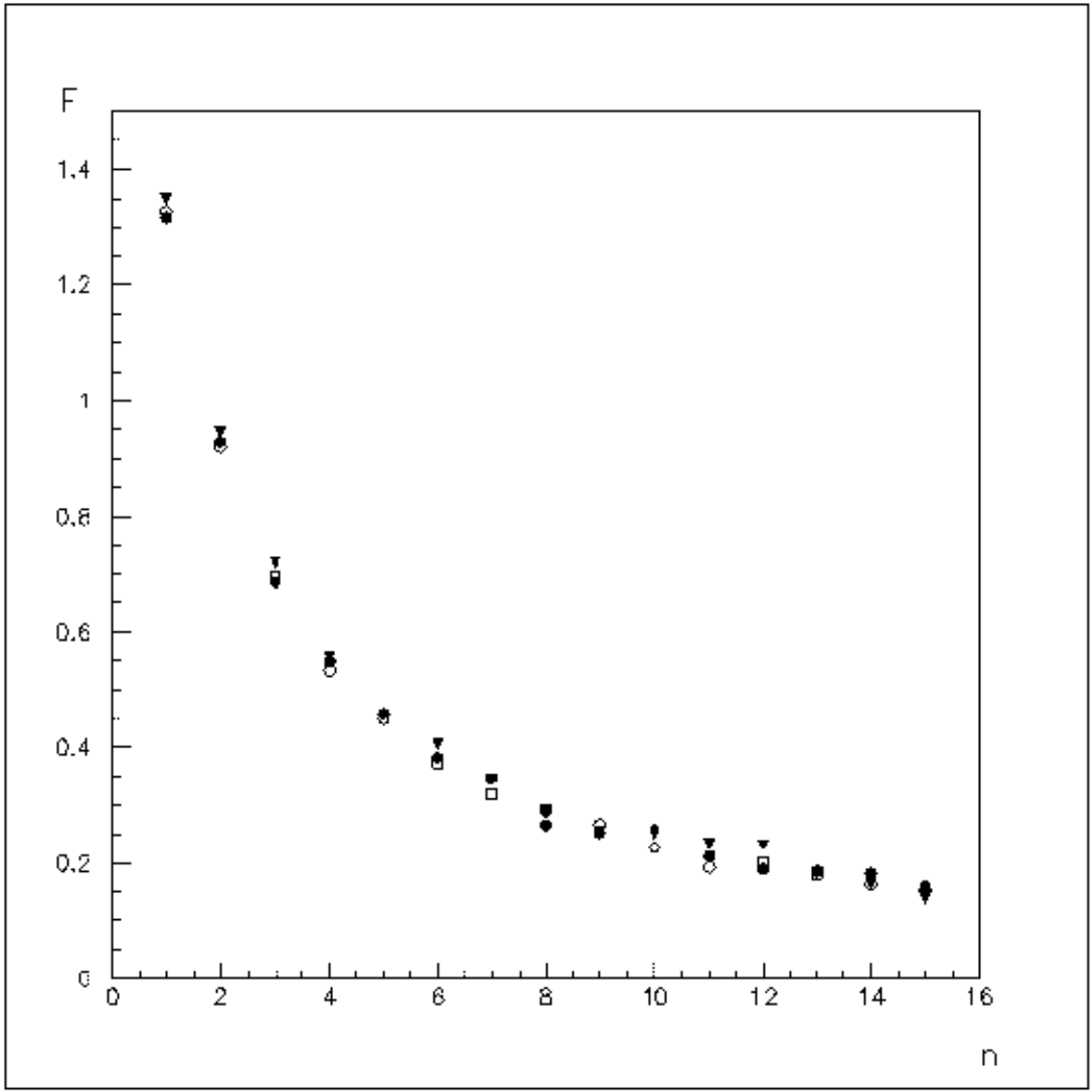

**Fig. 3** - Plot of the function $F(n)$ for different values of $\rho$.

This predicted behavior corresponds to the experimental findings, according to which there is: (1) a cylindrical shielding region, (2) without diffraction at its border.

In fact, the coherence domains inside the disk are very small on a macroscopic scale and a single critical region cannot be larger than one of these domains. Thus the total shielding factor is given by the contribution of several domains. (See Fig. 4. Also note that the effect of different domains can superpose over the thickness of the disk: Fig. 5.) The distance of the proof masses from the superconducting disk corresponds to a large value of $n$, such that $F(n)$ has already reached its constant asymptotic value. This explains the observed features (1) and (2) above.

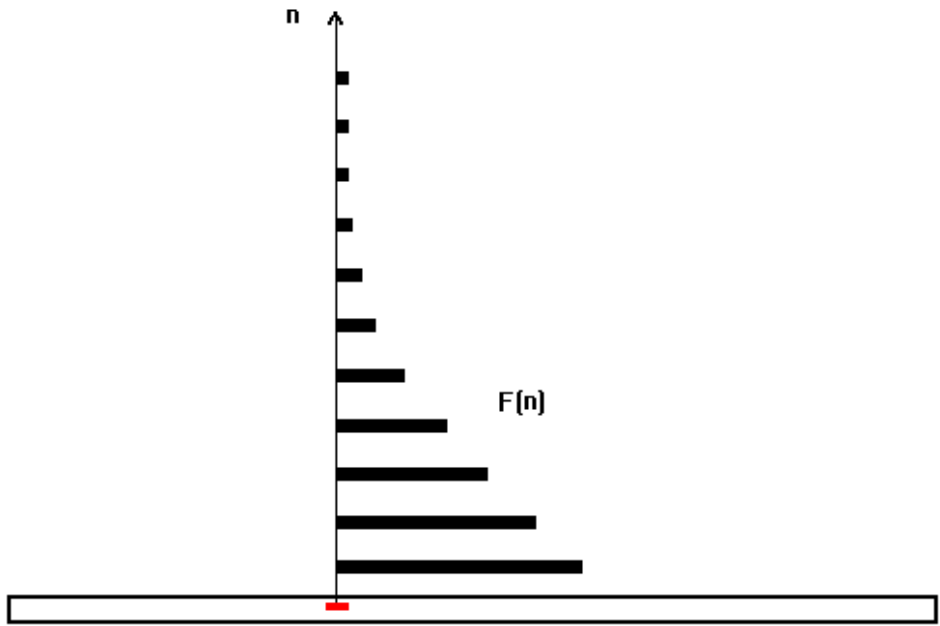

**Fig. 4**

This graph shows, for purely illustrative purposes, the behaviorof the shielding factor produced by a single critical region (coherence domain) along its vertical axis. The shielding factor is proportional to the adimensional function $F(n)$, where $n$ is the distance from the critical region in units of the region's radius. The present graph is in practice the same one as in Fig. 3, transformed into a bars diagram, rotated by 90 degrees and placed in space.

The coherence domain corresponds in our model to a region with over-critical density, and in our calculations we assumed it for simplicity to have the form a squeezed ellipsoid ($\rho\alpha>1$). It lies within the superconducting disk and is surrounded by similar domains.

The total macroscopic shielding factor is given by the sum of the effects of the single domains. Note that the size of the domain depicted in the figure is exaggerated as compared to the size of the disk. Since the parameter $n$ represents the distance *in units of the disk radius*, it is clear that the exponential decay of the shielding factor predicted by our model actually takes place very close to the disk.

What remains at macroscopic distances, and gives the observed effect, is the constant "tail" of the exponential, which amounts to approx. the 10% of its value for $n=1$ and does not vary with increasing distance.

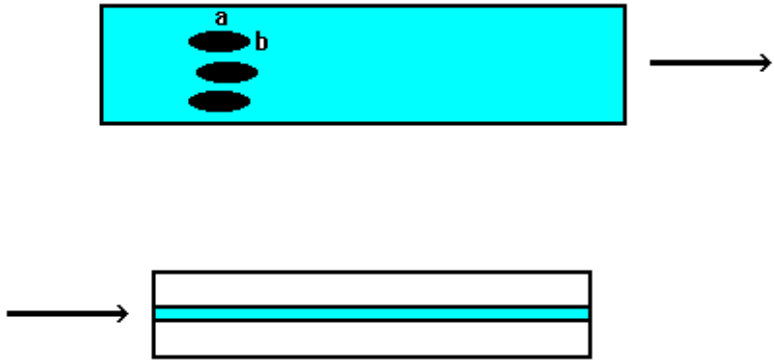

**Fig. 5**

The effects of more coherence domains, each of thickness $b$, are summed in such a way that the total shielding factor is proportional to the thickness of the layer of the disk where critical density can be reached. We recall (arrow) that this layer is quite thin compared to the whole disk (see Point 3: "Density pattern of the superconducting carriers in the composite disk").

The shielding produced by a single domain is proportional also to its width $a$; thus the total shielding factor is proportional to the average value of $a$ but not to the radius of the disk.

(Actually, a larger disk radius enhances the effect in an indirect way, because it means a larger tangential velocity during rotation andthis leads in turn to a thicker "critical layer" - see Point 3.)

## (3) Density distribution of the superconducting carriers in the composite disk.

This is a more "phenomenological" issue. While the Points 1-3 above were mainly concerned with the dynamics of the gravitational field, this point is concerned with the properties of the superconducting material.

It is important to know the density distribution of the condensate of the superconducting charge carriers in the ceramic disk, because this distribution defines the critical regions, where the density or its gradient exceed the threshold value.

In our earliest work (SHI) we limited ourselves to state that the fast rotation of the disk and the applied magnetic fields were probably responsible for strong variations in the condensate density. Here we push our analysis a bit further. It is not necessary, to this end, to make any special hypothesis about the microscopic mechanism of high-$T_c$ superconductivity. It is sufficient to consider an effective model, based upon an order parameter, like the Ginzburg-Landau (GL) theory.

In the GL theory the variations of the order parameter, and thus the variations of the condensate density, are always such that this density vanishes at the boundaries between superconducting and non superconducting regions. In our case, on the contrary, we are interested into a local *increase* of the density. How is it possible to achieve this?

All the phenomenology of the experiment shows that: (i) the two-phases structure of the superconducting disk is essential in order to obtain the shielding phenomenon; (ii) the effect takes place in conditions - fast rotation, high frequency fields applied - which are very far from the static limit, and thus cannot be adequately described by the GL theory.

Therefore it seems that the presence of an interface between two phases of the disk having different superconducting properties, combined with rapid rotation, is able to produce a local increase in the density which is not predicted by the GL theory and as such quite rare.

We do not have any proper formalism to describe this situation yet, namely a composite superconductor in rapid rotation. We can only resort at present to a qualitative model.

In our earlier work BOS we depicted the superconducting carriers in the upper part of the disk as a perfect fluid which flows, during rotation, close to a "border" represented by the lower part of the disk, which is non superconducting at the operation temperature. The relative sliding velocity, due to the fast rotation of the disk, is very large in comparison to the average velocities typical of the superconducting charge carriers in stationary conditions. This - we hypothesized - would produce a density increase near the border.

This phenomenological model is probably incorrect at the microscopic level, because the motion of the superconducting carriers cannot really be a "viscous" motion exibiting velocity gradients; moreover, the microscopic structure typical of any Type II superconductor must be taken into account. Nevertheless, the model has the merit to stress the importance of the interface between the two parts of the disk, and is also supported by several experimental observations, which give a clear feeling of a fluid-dynamical behavior of the system.

In particular, one observes that the shielding effect is considerably increased during the disk braking phase, when the relative sliding velocity of the superfluid in the upper part of the disk is much increased by virtue of the inertia of the fluid.

Furthermore, it has been observed since the very first experiments in 1992 that the effect remains even when, after reaching a rotation speed of 4000-5000 *rpm*, all lateral magnets are turned off and the disk rotates by inertia. As long as the disk rotates, the shielding effect persists. This confirms the purely kinematical role of rotation.

See the "Discussion" below for a closer analysis; further work is in progress.

## (4) Energetic balance and effective equations for the gravitational field.

Like for Point 3, also this point represents a wide phenomenological issue, which can be studied independently from the microscopic gravitational mechanism.

Several questions arise spontaneously here, also for the non-specialist. First of all, it is clear that any physical mechanism which expels from a region the gravitational field - or a small part of the field - must have some consequences on the global energetical balance. If a mechanical process inside the shielded region is affected, the total energy must be conserved.

The most frequent consequence of the shielding phenomenon seems to be an *increase* in the mechanical energy of the shielded objects (see BOS). This energy must come from an external "pumping" and if the latter is absent or inefficient, the shielding will be inhibited.

The whole phenomenology of Podkletnov's experiment indicates that the pumping is done by the high-frequency components of the applied magnetic field. This kind of pumping is familiar in atomic physics, for instance in lasers or in systems with magnetic resonance.

We believe that the high-frequency field allows the forming of regions with overcritical density of superconducting carriers, or it "activates" these regions and triggers the runaway of the field as described in Section 2b. It is a common experience that the runaway of an unstable system cannot take place unless a channel for the energetic exchange is available, in such a way that the total energy balance is respected (compare for instance phase transition phenomena).

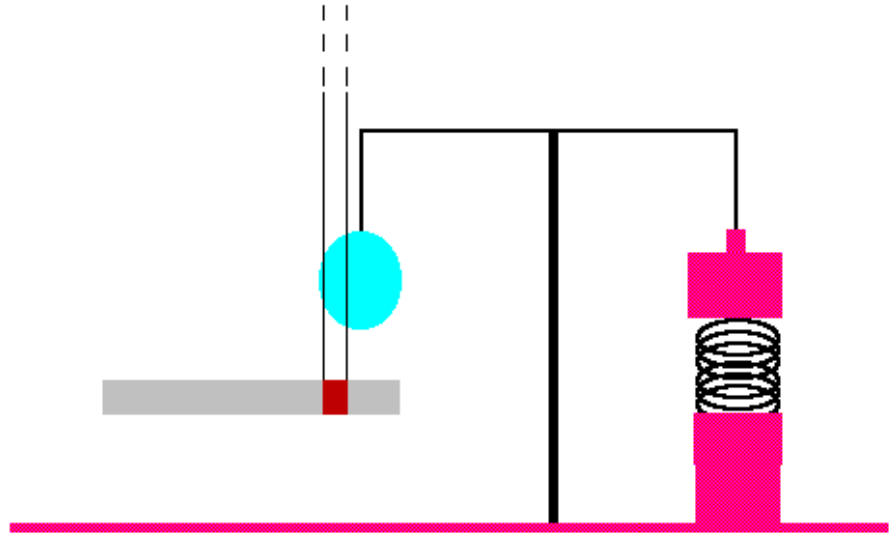

**Fig. 6** - Some energy pumping process is needed, for gravitational shielding to arise.

When in a region of the superconducting disk (red square in the figure) the density of the condensate exceeds the critical value, the gravitational field is virtually affected in that region. This produces a shielding cylinder: a proof mass placed above that region will become slightly lighter. However, this (still virtual) process can in turn produce mechanical energy! For instance, if there is a counterweight, we can compress a spring in this way; when the shielding ceases, the two masses will start to oscillate.

All this means that the state "region in the disk with overcritical density, shielding activated" has higher energy than the "normal" state. A transition to this state is possible only if energy is supplied from the outside. In Podkletnov's apparatus, this "activation" energy comes from the external high-frequency e.m. field.

This representation of the role of the high-frequency field is obviously approximated, and a complete theory should eliminate the distinction, useful but artful, between the shielding process, the pumping which activates it and its energetic consequences. From the causal point of view, the three phases are in the following order:

*Energy pumping from the outside* →

→ *Runaway of the field in the critical regions and shielding* →

→ *Changes in the mechanical energy of objects in the shielded region*

We could also put an arrow, however, which from the last point goes up to the first one, with a sort of *feedback*; this is because any variation in the mechanical energy of the objects inside the shielded region requires an energy supply, and in the absence of this the runaway of the field in the critical regions with subsequent shielding is inhibited.

Beside a direct interaction between the high frequency e.m. field and the superconducting carriers, indirect interactions of several kinds could be present. The e.m. field might first excite some mode peculiar of the crystal lattice of the ceramic material, or a mode belonging to the spectrum of the superconducting state; then the energy would be released for the shielding process. This could explain the resonant behavior of the system at certain frequencies.

Given our poor knowledge of the details of the pumping process, Podkletnov's strategy to send on the disk a strong and wide spectrum AC field appears to be drastic but somehow effective - at least as long as the superconductor is not heated too much.

It was early realized that the observed clear-cut cylindrical form of the shielding region and its vertical extension (at least a few meters) are very unusual and represent a puzzle from the conceptual point of view. For comparison, a metal shield placed in an electrostatic field produces in general a cone-like shielding region, with relevant border effects. On the other hand, if the SC disk would emit an hypothetical secondary field, or some radiation, this would probably appear like a divergent beam. (Also note that in this case one should expect some emission downwards, which does not appear to be the case.)

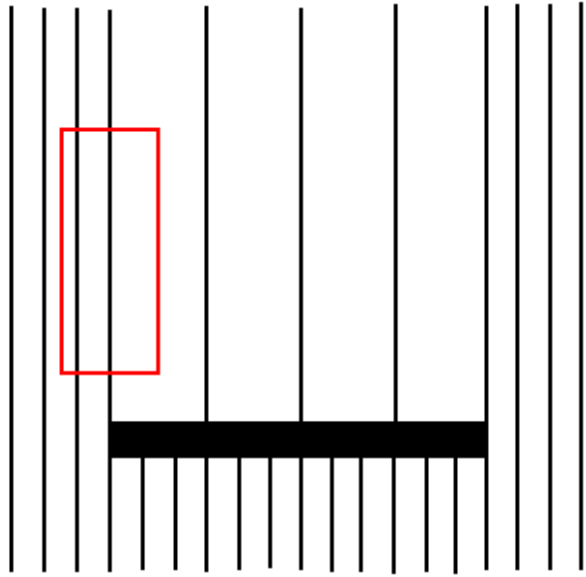

**Fig. 7** - Non-conservative character of the modified field.

The observed modified field pattern is clearly non conservative (Fig. 7). If a test mass makes a round trip going up inside the shielding cylinder and coming down outside, the gravitational field exerts a net work on the mass. Formally this amounts to say that the observed field is not the gradient of a gravitational potential. The Einstein equation for a weak static field still holds, namely

$$\text{div}\boldsymbol{\Gamma}_{00} = 0, \text{ with } a = \Gamma_{00}/m$$

($a$ is the acceleration of the test mass $m$, in the limit when the velocity of $m$ is much smaller than the light velocity); however, the associated equation

$$\boldsymbol{\Gamma}_{00} = \text{grad } h_{00}$$

is not true.

The non conservative character of the observed field configuration originates from the peculiar coupling of the gravitational field to the pumping e.m. field - coupling mediated by the condensate in the superconductor. The mechanical energy gained by a test mass along a round trip is furnished by the e.m. field and of course there is no global gain.

An interesting experimental observation in support of our interpretation is the following. If the energetic pumping is inefficient (that means, the applied e.m. field is too weak, or its frequency too low, or the test mass too heavy) one observes a weak "expulsive" force acting on the test mass at the border of the shielding cylinder. This force tends to restore the zero-work condition along a round trip.

## (5) The threshold problem.

We mentioned earlier the condition defining a critical region in the condensate of the superconducting carriers: a certain function $\mu^2(x)$ of the condensate density and gradient must be larger than the "natural" cosmological term $\Lambda/8\pi G$. Therefore this term represents a threshold value for the density.

The value of $\Lambda$ is not known a priori. There are some upper limits, deduced from astronomical observations, but also some indications that $\Lambda$ scales with the distance and its effective value is larger at small distances.

Do Podkletnov's data really suggest the existence of a threshold density? Or does the shielding effect seems rather to depend in a *continuum* way on the density? To give a proper answer, one must take into account the pumping process, because it could happen in some case that the condensate density is well above the threshold value but the effect does not take place due to absent or inefficient pumping.

This issue has been discussed in BOS. In order to conclude that a threshold exists, we must be in conditions of efficient pumping. If in these conditions the shielding effect is observed only above a certain local condensate density or density gradient, this is a point in favour of the treshold.

It was observed, for instance, that disks without two-layer structure, and thus without density gradients, do not produce any shielding effect. This is a point in favour of the existence of a treshold.

If a treshold exists and the pumping is efficient, then any increase or diminution in the strenght of the effect can be interpreted as following an increase or diminution of the number or size of the critical regions.

Note that with absent or inefficient pumping even a condensate whose density exceeds by far the threshold value fails to produce any gravitational anomaly. This is probably the case of superfluid helium, which is much more dense than an electronic condensate: since it is electrically neutral and its temperature is extremely low, it cannot be subjected to any pumping of reasonable efficiency.

# (1) *Discussion* - Relevance and innovation of the effect. Impossibility of an explanation within General Relativity.

According to General Relativity - our best present theory of gravity - the dynamics of the gravitational field and its coupling to the mass-energy-momentum density which generates it are described by the (classical) Einstein equations. These are non-linear partial differential equations involving the components of the metric tensor and its first and second derivatives. They are similar, under several respects, to Maxwell equations, though more complicated and non-linear.

In very simplified terms, we can say that Einstein equations allow to find the gravitational field as a *response to a source* - linear in a first approximation, or non-linear in the presence of strong mass-energy densities. The proportionality constant between field and source is of the order of the Newton constant $G$ for linear responses and even smaller, of the order of $G/c^n$, for non-linear responses. There exist static fields and fields propagating like waves, but in any case their strength is related to the mass of the source which has generated them.

The only sources close to us which are massive enough to generate a detectable field are the earth, the moon, the sun and, to a smaller extent, the other planets of the solar system. Any other object or physical system available on a laboratory scale, irrespective of its chemical composition or microscopic structure, generates gravitational fields of exceedingly small strength. These fields can be detected through very sensitive instruments, but they are typically of the order of $10^{-9}$ $g$ or less ($g \approx 9.8$ $m/s^2$ is the field generated by the earth at its surface).

These observations are well known and lead to the conclusion, in full agreement with Einstein equations, that the gravitational field generated by a very massive field is in practice unaffected by the presence of any other body whose mass is much smaller. In particular, is does not seem possible that the gravitational acceleration $g$ at the earth surface can be affected, through any human-sized apparatus, by more than approx. 1 part in a billion.

The conclusion above rests, as mentioned, upon the hypothesis that the equations of classical General Relativity are appropriate to the situation.

It is known that quantum mechanics brings in some very small corrections to the classical equations of any field, including the gravitational field. In the quantum view, the field oscillates in a approximately harmonic "potential"; these oscillations take place around a minimum value corresponding to the classical field strength.

Usually the quantum fluctuations are irrelevant on a macroscopic scale. One can show, however, that the presence in a region of space of coherent vacuum energy ("zero point energy") modifies the potential in which the gravitational field oscillates. Zero point energy is present in macroscopic systems - that means, systems well above the atomic scale - which are described as a whole by a single wave function. If the zero point energy term was present uniformly in all space, it would not bring any consequence: the gravitational field of the entire space would react exactly in such a way to reset the zero of energy. Things are different, however, if the zero point energy term is present only in a well-defined small region of space; in this case it produces a localized instability (see Point 2).

## (2a) *Discussion* - Coherent coupling between a Bose condensate and the gravitational field.

### Quantum nature of the effect. Differences with the classical case.

A Bose condensate is formally described by a classical field $\phi_0(x)$, like a sort of ideal fluid. We assume that the lagrangian of this field has the standard scalar form and that external conditions define the field density, so that $\phi_0$ behaves in the functional integral (or "quantum partition function") of the system like an external field and not like an integration variable.

At the classical level, in order to find the gravitational effects of $\phi_0$ we take the first variation of the action and obtain the Einstein equations with a small cosmological term originating from the coupling to the fluid. This gives an extremely small correction to the vacuum Einstein equations.

Let us check this classical result formally. The energy-momentum tensor of the condensate is

$$T_{\mu\nu} = \partial_\mu \phi_0^* \, \partial_\nu \phi_0 - g_{\mu\nu} L(\phi_0) =$$

$$= \partial_\mu \phi_0^* \, \partial_\nu \phi_0 - g_{\mu\nu} (1/2 \, \partial_\alpha \phi_0^* \, \partial_\alpha \phi_0 + 1/2 \, m^2 |\phi_0|^2) =$$

$$= \partial_\mu \phi_0^* \, \partial_\nu \phi_0 - g_{\mu\nu} (1/2 \, \mu^2)$$

The total lagrangian, including the so-called minimal coupling of the condensate with the gravitational field, is

$$L = L(g) + 8\pi G \, T_{\mu\nu} \, h_{\mu\nu};$$

$$T_{\mu\nu} \, h_{\mu\nu} = h_{\mu\nu} \, \partial_\mu \phi_0^* \, \partial_\nu \phi_0 - (\text{Tr } h)(1/2 \, \mu^2)$$

and the corresponding Einstein equations are

$$\delta L / \delta h_{\mu\nu} = R_{\mu\nu} - 1/2 \, R \, g_{\mu\nu} + 8\pi G \, T_{\mu\nu}$$

$$= R_{\mu\nu} - 1/2 \, R \, g_{\mu\nu} + 8\pi G \, (\partial_\mu \phi_0^* \, \partial_\nu \phi_0 - 1/2 \, g_{\mu\nu} \mu^2) = 0$$

The last term is of the form $g_{\mu\nu}\Lambda$, corresponding to a "local" cosmological constant. It is also interesting to take the trace of these equations. We obtain

$$R = 8\pi G \, \text{Tr } T = 8\pi G (-\partial_\mu \phi_0^* \, \partial_\mu \phi_0 - 2m^2 |\phi_0|^2)$$

where we see that the scalar curvature produced by the fluid is very small, due to the small value of $G$ (we work in units $c=1$).

In conclusion, the fluid generates a gravitational field which is approximately proportional to its mass-energy-momentum density. The main factor is the mass density distribution and the forces leading to this particular distribution do not play an important role. This clearly holds, in General Relativity, for any fluid, no matter if coherent or not. Equations like those above are employed in cosmological and astrophysical models, to study the internal structure of the stars etc.

Our attitude must be different if we consider the quantum nature of the gravitational field. In this case the field fluctuates and its dynamics is represented by a functional integral weighed by the factor $\exp(iS[g]/h_{Planck})$. This means that the field can assume any configuration $g$, with a probability proportional to this factor.

Clearly the preferred configurations are those near the stationary points of $S$, i.e. those obeying the classical Einstein equations. We know that the action has the form $S=\int\sqrt{\{\det g\}}R$ and a stationary point is given by $h=0$, $R=0$ (flat space). There are some configurations however, the so-called *zero modes*, for which $h$ and $R$ are not zero, but the integral vanishes and $S$ is zero. Therefore these configurations are as likely as flat space.

Why does the field usually prefers to be in the flat space configuration and not in a zero mode? Because there is in the action a small additional spacetime independent term, called the negative intrinsic cosmological term, which suppresses them, as can be seen by expanding the action to second order.

The existence of a negative intrinsic cosmological term has been demostrated by numerical simulations of Euclidean quantum gravity near equilibrium. Also independently from these simulations, we take it as one of the fundamental assumptions of our model (see Table II), and there are several indirect evidences of it.

As we saw above, expanding to second order the action of a condensate coupled to gravity one finds a positive cosmological term which cancels the intrinsic negative term and leads to an instability.

It is important to stress that in a quantum mechanical context the quantum coherence of the fluid described by $\phi_0$ is essential. An incoherent fluid is not represented by a field like $\phi_0$, but by an ensemble of pointlike particles whose energy-momentum tensor is of the form

$T_{\mu\nu} = \Sigma_i \int ds_i \, p_{i,\mu} \, p_{i,\nu} \, \delta(x-x_i)$

Thus an incoherent fluid is not able to cause any instability.

Finally we observe that also an electromagnetic field in a coherent state might play a role comparable to that of $\phi_0$. This point deserves further investigation, even though a magnitude order estimate shows that a coherent e.m. field cannot achieve the mass-energy density present in a Bose condensate.

**Meaning of $\phi_0$.**

In the works where the coherent coupling is discussed (SHI, LCC, BOS) the classical field $\phi_0(x)$ is introduced as the mean value of a quantum field: $\phi_0=<0|\phi(x)|0>$. This definition raises formal problems for the definition of the state $|0>$, because this should be at the same time a relativistic invariant state, and the ground state of a condensate of massive particles.

It is therefore more convenient to work just with the functional integral of the system, without mentioning the states. We assume that $\phi(x)=\phi_0(x)+\phi'(x)$, with $\phi_0(x)$ a *relativistic* (this is necessary for the gravitational coupling) classical field describing the condensate.

While $\phi'$ is a quantum field, $\phi_0$ is an external classical field, and not an integration variable in the functional integral. We assume this to be a natural description of the condensate in the present context. We shall identify better $\phi_0$ a posteriori; in particular, we are interested in its relation to the density of superconducting charge carriers.

The relativistic hamiltonian, or energy density, of this field is equal to $\mu^2(x)$ as given in eq. (1). This hamiltonian does not take into account the external fields (mainly the magnetic field) and the boundary conditions (normal regions of the superconductor) which let $\phi_0(x)$ take its particular value at every point $x$. It just accounts for the mechanical energy of the fluid: the rest energy + the term due to the 4-gradient (which reduces in practice to the spatial gradient, because the time derivative is divided by $c$).

Since in natural units the mass of the charge carriers is of the order of $10^{10}$ $cm^{-1}$ and usually the spatial variations of $\phi_0$ take place on a scale much larger than $10^{-10}$ $cm$, we can at first disregard the gradient term. Then, consistently, we can estimate $\mu^2 \approx mN \approx m^2V|\phi_0|^2$ (compare BOS), from which we get the relation between $\phi_0$ and the energy density. If the coherence length of the superconductor is very small, it may be necessary to introduce a small correction to this relation, but the magnitude order of the total energy will be basically unaffected, and the same holds for $|\phi_0|$.

## (2b) *Discussion* - Induced instability and "runaway" of zero-modes.

**"Pinning" or "runaway" of the field?**

In Ref. BOS we mentioned a "pinning" of the gravitational field within the critical regions. Here we speak instead of a "runaway" of the field. The two terms are basically equivalent, since both refer to the behavior of the field in a potential where the $h=0$ value is unstable and therefore the field runs away from this value and gets pinned - almost independently from the initial conditions - at a strength $h'$ different from zero. $h'$ is determined by terms in the potential higher than second order.

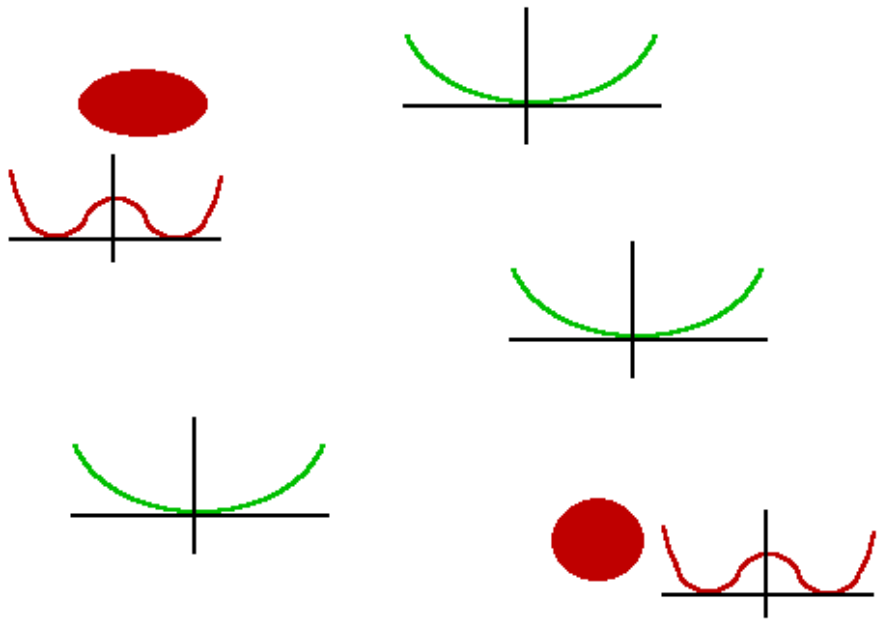

**Fig. 8 -** Double-well potential for the field, localized to the critical regions (red).

The most typical example of this kind of potential is the double well potential (Fig. 8). Actually, in the Ref. GAU we used a potential of this kind in order to compute the effects of a local pinning on the field propagation in all space. That computation describes *exactly* the case of a scalar massless field with local pinning. We recall that the correction to the propagator is proportional in that case to the parameter $\gamma=\xi f^2$, in turn proportional to $\mu^2$ - the local imaginary mass term.

On the other hand in the gravitational case, even though we have a term $\mu^2$ (the "positive cosmological term") in the lagrangian comparable to the imaginary mass term, it is not the value of $\mu^2$ which enters directly in the correction to the propagator. In other words, $\gamma$ is not proportional to $\mu^2$. If it was so, than the shielding effect would be irrelevant, because $\mu^2$ is extremely small.

We are not able, in the gravitational case, to relate directly $\mu^2$ (which depends on the condensate density in the critical regions) to $\gamma$ (which is evaluated experimentally, through eq. (2), from the observed shielding strength). This is because the instability induced by the $\mu^2$ term in the gravitational lagrangian cannot be compared with that due to a double well potential in the scalar case; it it much worse, since there exist gravitational modes for which the kinetic term in the action - usually stabilizing against local variations, as it contains a gradient squared - is not effective.

For this reason we believe that the exact value of $\mu^2$ is not important, as long as it is larger than the threshold value $|\Lambda|/8\pi G$ and can thus trigger the instability. It follows that the computation in Ref. GAU, exactly valid for a scalar field, represents in the gravitational case only a useful model for a pinning of the field following an instability, while the parameter $\gamma$ must be fitted from the experimental data. As explained in Section 2b, $\gamma$ is the sum of the products of the unknown runaway probabilities $\xi_i$ by the unknown strengths $h_i^2$.

**In which sense it is possible to use the Euclidean formalism.**

It is known that the Euclidean Einstein action is not bounded from below due to the so-called "conformal modes". The use of the Euclidean formalism to study the stability of our system may therefore appear quite arbitrary. Consider, however, the following points.

(i) The phenomenon under investigation is essentially macroscopic. Its typical distance and momentum scale is such that terms in the action of the order of $\delta R$ can be disregarded. Equivalently, it is possible to insert in the action a cut-off on the momenta which eliminates the conformal modes.

(ii) The Euclidean formalism is employed only in a perturbative context, as the analytical continuation of Minkowski space + small fluctuations. We use the Euclidean formalism in order to ensure the positivity of the function $\mu^2(x)$ (see 2a) and in order to take advantage of some known techniques for the computation of the static potential (see 2c).

Furthermore, we use the Euclidean formulation in order to exhibit the stabilizing effect of a negative cosmological term and the de-stabilizing effect of a positive comological term, but still in the context of a weak field approximation. The same results can be obtained in the Minkowskian theory, where a positive comological term corresponds to an imaginary mass.

After realizing that an instability arises, we do not try to describe the behavior of the field through the Euclidean formalism; we just suppose that the runaway process stops at some field strength, in such a way that the strengths $h_i$ have certain unknown probabilities $\xi_i$ (see 2b).

## (2c) *Discussion* - Constraint on the gravitational field and "tunneling".

### Why the shielding region has a cylindrical shape.

Intuitively, one would expect the shielding region above the superconducting disk to have the shape of cone, instead of a cylinder (see also Fig. 9). The observed cylindrical shape implies that: (i) the source of the field of the earth is seen as pointlike; (ii) there isn't any kind of "diffraction" at the disk border.

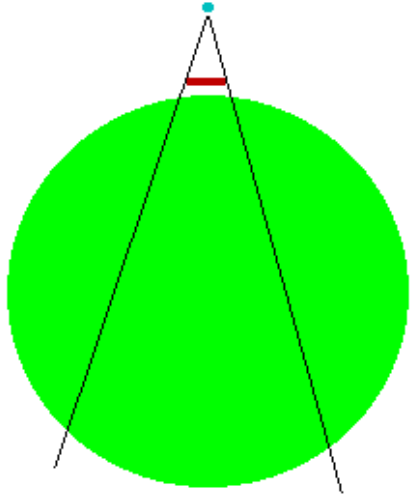

**Fig. 9** - Why is the shielding region cylinder-like instead of cone-like?

This is not easy to explain. Intuition would suggest a cone-like shielding region above the superconducting disk (the basis of the cone being on the disk), instead of the infinite cylinder-like region observed by Podkletnov. In fact, when the elevation of the proof mass above the disk increases, the proof mass apparently "sees shielded" only a part of the Earth. This geometrical factor can be computed exactly (see G. Modanese, supr-con/9601001).

Nevertheless, it turns out that when computing the shielding factor for the Newtonian part of the gravitational force (that with dependence $1/r^2$), the mass of the Earth can still be regarded, as usual in gravity problems, as concentrated in the center.

Condition (ii) is consistent with our theoretical model, as shown by the computation in GAU, under the reasonable hypothesis that the coherence domains in the superconducting disk are very small in comparison to the disk size.

Condition (i) is a general consequence of our theory, too. In fact, in order to compute the static potential energy of the interaction of two bodies with masses $M$ and $m$ one adds to the gravitational action $S_g$ the actions $S_M$ and $S_m$ of the two bodies and then computes the quantum average $<\exp(S_M+S_m)>_g$. For pointlike bodies, the action is simply given by $\int ds$, where the invariant interval $ds$ computed along the trajectory of the body, which is a geodesic line in any field configuration $g$.

For an extended body, the center of mass (CM) still follows a geodesic line, but in addition to the line integral we must also insert a volume integration over the parts of the body: $S=\int dV\rho(\mathbf{x})\int ds(\mathbf{x})$. This implies in practice that in addition to the minimal gravitational coupling, of the form $(h_{\mu\nu}v_{CM,\mu}v_{CM,\nu})$, there is a coupling with the *derivatives* of $h$ and quadrupole, octupole etc. terms of the form $(\partial_\alpha\partial_\beta h_{\mu\nu}Q_{\alpha\beta\mu\nu})$, etc.

These terms do give some contributions to the static interaction potential, when inserted into the quantum average $<\exp(S_M+S_m)>_g$, but these contributions depend on the distance $r$ as $r^{-2}$, $r^{-3}$, etc. So if we keep the multipole terms into account when computing the "shielding correction", we find that the correction to the Newtonian term $U\approx r^{-1}$ depends only on the coordinates of the center of mass. In other words we can regard the two bodies as pointlike, as long as we are interested only in the Newtonian potential, which is usually by far the dominant part of the interaction.

**What is the analogue of $f_i$ in the gravitational case?**

For a scalar field $f$ one can impose a local constraint (see GAU) through a double well potential of the form $U(f)\approx[f^2(x)-f_i^2]$. (More precisely, the r.h.s. is multiplied by a function $f_\Omega(x)$ which has support in the critical region $\Omega$.) In the gravitational case we could write instead $U(h)\approx[h_{00}^2(x)-h_{00,i}^2]$, but in principle $h_{00,i}$ cannot be a constant, because the zero-mode, which is the new minimum of the action in the presence of instability, depends on $x$.

Since there are several zero-modes close to the minimum of the action, we could take an average of these modes with respect to $x$ and obtain a constant field. It is better, however, to consider the field at one single point: here the field takes values $h_i$, with probabilities $\xi_i$, and the same holds for any point, with the same $h_i$'s and $\xi_i$'s; the only difference is that for points close to the border of the critical region $\Omega_i$, $f_\Omega(x)$ is not equal to 1, but goes gradually to zero. Thus we have a constraint of the form

$$\sum_i \xi_i \int_i dx\, f_\Omega(x)\, [h_{00}^2(x)-h_{00,i}^2]^2$$

Actually, it would be impossible to assign different probabilities to different points, except for those at the border of the critical region.

One easily checks that the effect of the change above on the final formula for the shielding factor is just to replace the parameter $\gamma$ with another unknown factor.

---

Table II - Unproved assumptions of our model.

(i) The Euclidean theory near equilibrium can be applied to quantum gravity.

(ii) The instability leads to a constraint, with probability $\xi_i$ for the field strength $h_i$.

(iii) There exists a small intrinsic negative $\Lambda$ and thus a threshold.

## (3) Discussion - Density distribution of the superconducting carriers in the composite disk.

In our earlier work we hypothesized that a local increase in the superconducting carriers density could be obtained in the composite disk in non-static conditions, due to the very fast flow of the "superfluid" in the upper part of the disk near the border of the lower (non superconducting) part. (Fig. 10)

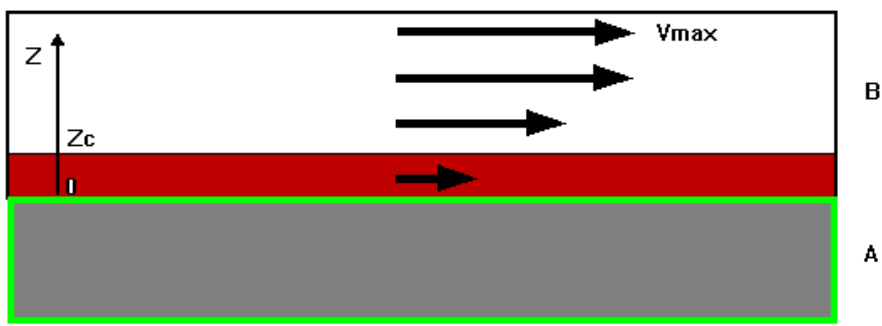

**Fig. 10** - Fluid dynamical model for the composite disk.

In fact, admitting that the relative drift velocity goes to zero at the border, one concludes from simple fluid dynamical arguments that the fluid density must *increase* at the border - the opposite of what usually happens in static conditions at any normal/superconducting interface. More precisely, the carriers density would reach a peak near the interface and then go rapidly to zero at the interface.

A velocity gradient near a flow boundary is the typical signature of a viscous fluid however, and this cannot be the case of the supercurrents. How could it then be possible to justify a velocity gradient?

We could regard the velocity as a mean velocity, and keep into account the "strong Type II" nature of the ceramic superconductor as well as the role of defects and impurities. These are much more frequent near the interface, and the latter is not clear-cut, being obtained through a thermal process which melts the upper part of the disk.

The mean free path of the superconducting carriers could be very long in the upper part of the disk and become gradually shorter towards the bottom of the disk. Therefore the superfluid would slow down at the interface - in the sense of the average motion - due to the interaction with obstacles and impurities in the lattice.

This could also explain why the disk tends to heat up just when the largest values of the shielding factor are observed: the reason would be that the almost-resistive behavior depicted above is just what is needed to achieve a local increase in the density of superconducting carriers.

Or perhaps the causes of the heat production are different (for instance, related to the pumping process), and a "mean path" of the superconducting carriers is meaningless in this context? All these questions are still unsolved; as we pointed out above, a proper theoretical treatment of composite Type II superconductors (not mentioning the peculiar properties of HTCs...) subjected to fast rotation and high-frequency e.m. fields is not available yet.

Another useful observation could be the following. We can say that the composite disk exhibits, in a certain sense, a "Tc gradient", from the top to the bottom, from 92 K to approx. 60 K. The shielding effect was observed by Podkletnov slightly below 70 K.

Remember that the distance between the vortex cores in Type II material tends to zero when the temperature approaches Tc (Fig. 11). It is then clear that if a supercurrent flowing in the "good" part of the disk in deviated to the lower part, where there is less space available for the superconducting carriers, the density of the latter can increase even if their velocity keeps constant. The deviation can be caused by the external variable magnetic field, or by fluctuations and defects in the lattice. The density increase would in any case also be connected to the relative drift velocity of the carriers with respect to the rotating lattice.

This possibility is currently under investigation.

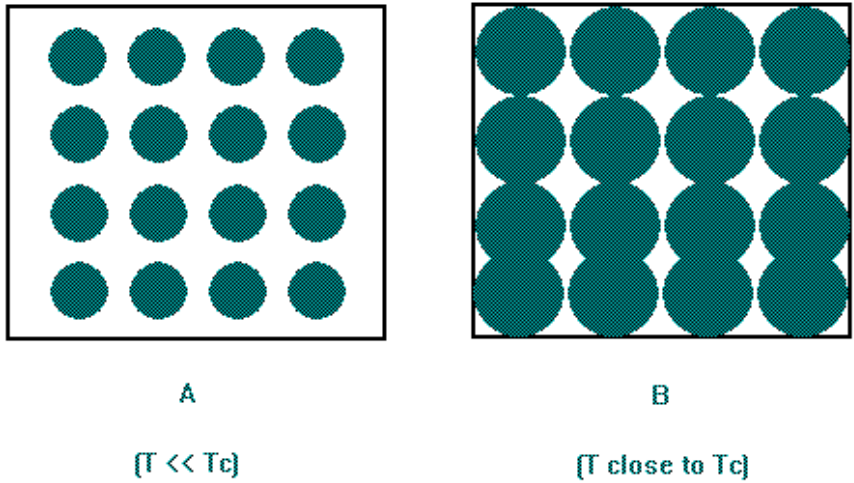

**Fig. 11** - Size of the flux cores in dependence on the temperature.

## (4) Discussion - Energetic balance, non conservative character of the field and equation $\partial_i\Gamma_i=0$.

As we saw in the first part, it is possible to keep into account the non conservative character of the gravitational field in the context of a phenomenological theory, in which the external energy source has been included and therefore we do not expect that the static force field is still equal to the gradient of the potential $h_{00}$.

We still expect, however, that the minimum of the action outside the superconductor corresponds to the Einstein equations; it follows, for weak fields, the equation $\partial_i\Gamma_{00}^{i}=0$, plus suitable boundary conditions. This admits as a solution the observed cylinder-like shielding region.

It is natural for a general-relativist to ask, at this point: what happens to a light ray when it crosses the shielding region? Is it deviated? And will a precision clock show any gravitational slow-down when it is first placed into the shielding region and then pulled out?

These phenomena are typically studied using the metric tensor. We know that in the presence of shielding the relation between the static force $\Gamma_i$ and the gradient of $h_{00}$ is spoiled. The question is: is it still possible to use the metric tensor to predict the trajectory of a light ray or the relation between the proper time of a frequency standard and the coordinate time?

These effects could also be tested experimentally, at least in principle. Light propagation across the shielding region could be studied, for instance, through interference measurements on a laser beam after several reflections at both sides of the region. It can also be shown (details will appear elsewhere) that the modification of the metric tensor inside the shielding region may lead to a shift in the frequency of a clock of about 1 part in $10^{11}$ - approx. 2 orders of magnitude larger than the precision of atomic clocks.

But apart from the several practical difficulties connected to these measurements, it is not clear yet on the theoretical side whether the gravity anomaly affects in the same way all the components of the metric, and whether it affects in the same way the static behavior of the field and its variations on a short time scale. Much work is on the way in this direction.

Another important issue is the compatibility between the shielding phenomenon and the equivalence principle (see Fig. 12).

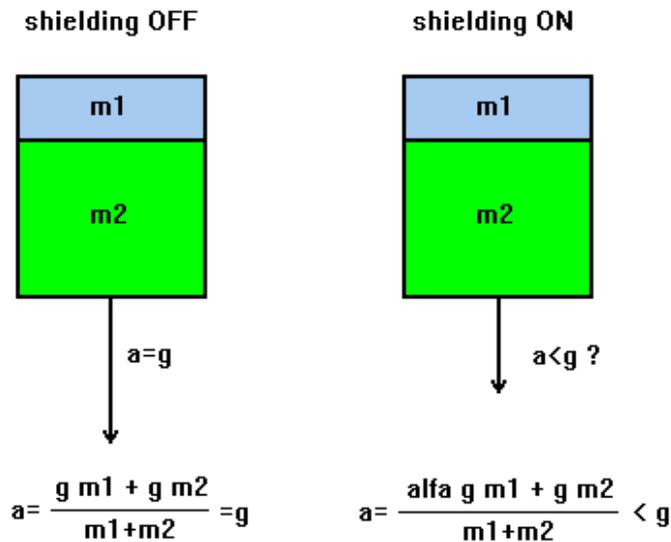

**Fig. 12** - A "flying machine" based upon the gravity shielding effect would violate the equivalence principle.

Imagine a box divided in two sections 1 and 2. Suppose that the lower part of the box, with mass $m_2$, contains a shielding apparatus, complete with power supply generator and everything. Now let the box be in free fall. If "the shielding is OFF", the acceleration of the box is equal to $g$.

Then you "turn ON" the shielding, say with efficiency $\alpha$; this means that the gravitational force felt by the mass $m_1$ over the apparatus is multiplied by a factor $\alpha<1$ (for instance, $\alpha=0.98$). Let us admit that the weight of $m_2$ itself is not affected.

It is easy to see that in this case the acceleration of the box becomes less than $g$. This is actually what desired, if we aim at building a flying machine. It means, however, that the gravitational mass and the inertial mass of the box are not equal, any more. And this represents a violation of the equivalence principle.

Note that the box is supposed to be isolated from the enviroment: it does not expel any jet of air or gas, nor it interacts with any external electric field, etc. In these conditions of free fall, one observer inside the box should experience total absence of gravity. He doesn't, however, if the shielding is ON. He feels some gravity, because its acceleration is lower than $g$. This, again, shows that the equivalence principle is violated.

If we do not accept the possibility of such a violation, we must admit that the shielding effect does not work like this. We must admit that if the shielding apparatus is rigidly connected to the Earth, then there is effective weight reduction of the samples suspended over the apparatus; but if the whole shielding apparatus is in free fall, then a reaction force from the samples on the apparatus arises, which makes the total weight variation vanish.

This means of course that it is impossible to build a flying machine using the gravity shielding effect. It is still possible however, in principle, to build a "lift".

**Frequency spectrum of the transient effect.**

It has been reported that the amplitude of the transient effect described by J. Schnurer appears to depend much on the method employed for its detection. This led us to the following conclusion.

(1) Probably the transient weight variation at the superconducting transition has its own characteristic frequency spectrum. This is because it is short-lasting, which brings in a frequency spectrum of course, and also because there could be some kind of temporal spikes in the effect.

(2) The frequency spectrum of the effect should match well the proper frequency (or frequencies) of the detector, in order to cause a good response in the detector. Things go like in a resonance effect.

So, roughly speaking, one should not figure the measurement as "the effect comes, the detector feels a transient weight diminution and displays this, then the effect goes and the detector ceases to display the weight diminution". It could be really so, only if the duration of the effect was of 30 seconds or more, and without any spikes.

The correct picture would be: the effect comes and acts like a force which perturbates the proof mass with a certain frequency spectrum. If the proper frequency of the mechanical system holding the mass is close to the typical frequencies of the effect, then an oscillation starts; if not, almost nothing happens, because the mechanical system holding the mass is either to rigid or too loose (compare also Fig. 6).

Probably the wooden arm used by Schnurer in his measurements or some other element in the arrangement has a certain vibration frequency which resonates with the effect. It would be very important to clarify this point.

While doing this hypothesis, we are figuring that in fact no pure weight diminutions were seen, but weight oscillations in both directions, due to this kind of resonance. Several balances take a few seconds to give an accurate response, and are therefore not suitable for measuring a weight which is supposed to vary quickly - or they start oscillating if the weight oscillates, but in some strange and unpredictable way. As a matter of fact, people who build balances never consider the possibility that weight can vary with high frequency, due to some little "kicks" coming from the bottom...